\documentclass[twocolumn,twoside]{revtex4-2}
\usepackage{graphicx}
\usepackage{fancyhdr}
\usepackage{relsize}
\usepackage[export]{adjustbox}
\usepackage[utf8]{inputenc}
\usepackage[]{hyperref}
\DeclareUnicodeCharacter{00A0}{ }
\usepackage[T1,T2A]{fontenc}
\usepackage[dvipsnames]{xcolor}
\usepackage{amsmath}
\usepackage{amssymb}
\usepackage{comment}
\usepackage{subcaption}
\hypersetup{
colorlinks,
linkcolor={blue},
linktocpage=true,
citecolor={red},
urlcolor={olive}
}

\begin{document}

\title{Experimental Searches For Heavy Neutral Leptons}

%
\def\babar{\mbox{\slshape B\kern-0.1em{\smaller A}\kern-0.1em
    B\kern-0.1em{\smaller A\kern-0.2em R}}}
\author{Sophie C. Middleton}
\affiliation{California Institute of Technology, Pasadena, California 91125, USA }

\begin{abstract}
\begin{center}
    \textbf{on behalf of the \babar~Collaboration}
\end{center}

The highly successful Standard Model is not complete. It does not explain the baryonic asymmetry in the Universe, the existence of dark matter or the non-zero masses of the neutrinos. Extensions of the Standard Model that propose the existence of additional Heavy Neutral Leptons (HNLs) are well-motivated and can explain several of these phenomena. In addition, light sterile neutrinos of $\mathcal{O}(\text{eV}/c^{2})$ can explain experimentally observed oscillation anomalies. The Neutrino Minimal Standard Model proposes HNLs with masses $\mathcal{O}(\text{keV}/c^{2} - \text{GeV}/c^{2})$, while more exotic models predict very large masses, up to the GUT scale.  Due to the multitude of models which hypothesize HNLs, the mass range to be explored by experiments is large. Experimental searches for HNLs can be conducted at existing neutrino, beam dump and collider-based experiments, and, depending on the signature, can constrain mixing between additional neutrinos and any of the three active neutrinos.


\end{abstract}

\maketitle

\thispagestyle{fancy}


\section{Motivations}

Heavy Neutral Leptons (HNLs) are predicted by many extensions to the Standard Model (SM). They possess mass and therefore interact via gravity, but have no electric charge, no weak hyper-charge, no weak isospin, and no color charge. They are singlets under all gauge interactions and are subsequently referred to as ``sterile neutrinos." Sterile neutrinos have long been used to explain the apparent smallness of the SM neutrino masses \cite{small}. HNLs interact only with the active neutrinos via mixing, and possibly the Higgs boson. Current experimental limits are outlined in Refs. \cite{PBC,Abdullahi:2022jlv}. The following sub-sections document how HNLs can help resolve several known issues with the SM and how they are ubiquitous in models that explain neutrino mass.

\subsection{Inconsistencies between Standard Model and Observation}

Standard Model predictions have been extensively tested and are found to be in good agreement with experimental results; in some cases with extraordinary precision of up to 1 part in a trillion \cite{PhysRevLett.100.120801}.  Nevertheless, there remains a need to extend the SM to explain several observational phenomena including: the baryon asymmetry in the Universe (BAU), the existence of dark matter, and the non-zero mass of the neutrinos.

The Neutrino Minimal Standard Model ($\nu$-MSM) \cite{neut_min} is one extension which requires the existence of three HNLs. It is capable of explaining the origins of neutrino masses, dark matter \cite{DM1} and the BAU \cite{DM2,astro}. The HNLs have Majorana masses below the electroweak scale and realise the see-saw mechanism with SM neutrinos. Two of the HNLs have masses in the $\mathcal{O}$(MeV/$c^{2}$ - GeV/$c^{2})$ range and a third, the dark matter candidate, has mass of $\mathcal{O}$(keV/$c^{2}$). $\nu$-MSM  is compatible with all current measurements. Models with GeV/$c^{2}$ scale HNLs are the subject of intensive theoretical study.  Heavy Neutral Leptons with a masses from $\mathcal{O}(100$ MeV/$c^{2}$) up to a few $\mathcal{O}($GeV/$c^{2}$) can be produced in decays of SM particles, while heavier states, of a few $\mathcal{O}($GeV/$c^{2}$), can be directly produced at colliders.

    \subsection{Incorporating Neutrino Mass into the Standard Model}

 Observation of neutrino oscillations has established the non-zero mass of at least two of the SM neutrinos. Since the discovery of neutrino oscillations a global experimental program has measured, with precision, most of the oscillation parameters using neutrinos from solar, accelerator, reactor and atmospheric origins. Absolute values of these masses are yet to be determined, but experiments have measured the mass squared differences, with current bounds detailed in Ref.\cite{PDG2020}.  It appears that mixing in the lepton-sector has a very different structure compared to that of the quarks. There is no current explanation to why that might be. In addition, there are still several remaining questions regarding neutrinos such as: establishing whether neutrinos are Dirac or Majorana particles, their absolute masses and whether they exhibit CP violation. Accurately answering these questions, and accounting for neutrino masses in a consistent framework, is the focus of global experimental and theoretical effort. 

The most intuitive way to extend the SM to include the observed neutrino masses would be to add a term coupling neutrinos to the Higgs field, analogous to that for charged leptons:

\begin{equation} Y^{\nu}_{\alpha \beta}\bar{L}^{\alpha} \cdot \Tilde{H}\nu^{\beta}_{R} + h.c.,\end{equation}
where $\Tilde{H} = i\sigma_{2} H^{*}$; $\sigma_{2}$ is the second Pauli matrix; $H$ is the Higgs field; $Y^{\nu}_{\alpha \beta}$ describes the couplings of the neutrino flavor states to the Higgs field; $L^{\alpha}$ is a $SU(2)_{L}$ doublet of left-handed leptons; and $\alpha, \beta$ are flavor indices. Incorporating neutrino mass in this way requires the Yukawa coupling to be
orders of magnitude smaller than the that for charged leptons and that right-handed neutrinos have no Majorana mass, despite there being no symmetry preventing it. In addition, the masses and mixing angles would be expected to have a similar hierarchy as for quarks - which is not the case.

The SM admits a dimension-five operator, the Weinberg operator, that is gauge invariant:

\begin{equation}
    \mathcal{L}_{5} = \frac{c^{[5]}}{\Lambda} L^{T} \cdot \Tilde{H}^{*}C^{\dagger}\Tilde{H}^{\dagger} \cdot L + h.c.
\end{equation}
where $\Lambda$ is the scale at which the particles responsible for lepton number violation become relevant degrees of freedom;  $c^{[5]}$
is a flavor-dependent Wilson coefficient; and $C$ is the charge-conjugation matrix.


The Weinberg operator leads to Majorana masses of the neutrinos after electroweak symmetry breaking. If the neutrinos are in fact Majorana particles, there are a multitude of models which can generate Majorana mass terms for left-handed fermions below the electroweak symmetry breaking scale. These models are collectively referred to as ``See-Saw Models" and can also account for the smallness of the neutrino masses without introducing an extremely small Yukawa coupling, they are categorized in the three ``types":

\begin{enumerate}
    \item \textbf{Type I Seesaw} \cite{Yanagida:1979as,Mohapatra:1979ia,Gell-Mann:1979vob,Schechter:1980gr} with a singlet fermion;
    \item \textbf{Type II Seesaw} \cite{Schechter:1980gr,Abela:1981nf,Minehart:1981fv,Cheng:1980qt} with heavy triplet scalars;
    \item \textbf{Type III Seesaw} \cite{Lazarides:1980nt,Mohapatra:1980yp} with triplet fermions.
\end{enumerate}

One common feature of models that explain neutrino masses is the existence of a new HNL state. Severe constraints exist for eV-scale Seesaw \cite{Canetti_2010} from cosmic surveys and Big-Bang Nucleo-synthesis (BBN) \cite{BBN}. However, more natural solutions remain at the GeV or TeV scale. For a search at the GeV-scale the Yukawa coupling is of $\mathcal{O}(10^{-5})$ and can be searched for at existing experiments. At the TeV-scale direct searches become less effective, since the Yukawa coupling is smaller  ($\mathcal{O}(10^{-6})$).

\subsection{Suggestions of Additional Sterile Neutrino States}


The existence of light sterile fermions with masses $\mathcal{O}(\text{eV}/c^{2})$ can also provide an explanation for observed anomalies in very short baseline oscillation measurements and cosmological data analyses \cite{sterile_lowmass}. Re-analysis of data from the GALLEX \cite{Kaether:2010ag} and SAGE \cite{galium} solar neutrino experiments has exposed an unexplained 14$\pm$5$\%$ deficit in the number of recorded $\nu_{e}$ - referred to as the ``Gallium anomaly" . In addition, numerous analyses of the flux of $\bar{\nu}_{e}$ from reactors have suggested a deficit of $\bar{\nu}_{e}$ at the 98.6$\%$ C.L. \cite{reactor} - denoted as the ``reactor anti-neutrino anomaly." When combined, both anomalies disfavour the no-oscillation hypothesis at 99.97$\%$ ($3.6 \sigma$). A third anomaly  - the ``accelerator anomaly" - stems from measurements at the LSND \cite{PhysRevD.64.112007} experiment which evaluated the oscillation $\nu_{e} \rightarrow \nu_{\mu}$ at a baseline of $L$ = 30m. LSND measured an excess of neutrinos at the level of 3.8$\sigma$ which could be explained by the existence of a sterile neutrino with a mass $\mathcal{O}(\text{eV}/c^{2})$. Further support for this excess was presented by the MiniBooNE experiment at the 2.8$\sigma$ level \cite{Mboone}.

\section{Searching for Heavy Neutral Lepton States}

Mixing between the beyond SM (BSM) heavy neutrino mass eigenstate and the active neutrino states can be parameterized by the extended Pontecorvo Maki Nakagawa Sakata (PMNS) matrix. The additional elements, $U_{l,n}$,  represent the mixing strength between the active neutrino flavor state, $l$, and the BSM $n$-th neutrino mass state:

\begin{equation}
    \begin{pmatrix}
    \nu_{e}\\
    \nu_{\mu}\\
    \nu_{\tau}\\
    \nu_{\text{\tiny L}}\\
    \vdots
    \end{pmatrix}
    =
    \begin{pmatrix}
    U_{e1} &  U_{e2} &  U_{e3} &  U_{e4} & \cdots\\
    U_{\mu 1} &  U_{\mu 2} &  U_{\mu 3} &  U_{\mu 4}& \cdots\\
    U_{\tau 1} &  U_{\tau 2} &  U_{\tau 3} &  U_{\tau 4}& \cdots\\
    U_{\text{\tiny L} 1} & U_{\text{\tiny L} 2} & U_{\text{\tiny L} 3} & U_{\text{\tiny L} 4} & \cdots\\
    \vdots &\vdots &\vdots & \vdots & \ddots\\
    \end{pmatrix}
    \begin{pmatrix}
    \nu_{1}\\
    \nu_{2}\\
    \nu_{3}\\
    \nu_{4}\\
    \vdots
    \end{pmatrix},
\end{equation}
where $L$ represents some hypothetical additional lepton flavor. Given that analyses of cosmological data and measurements of $Z$ boson decays, summarized in Ref.\cite{PDG2020}, are consistent with there being only three charged lepton flavors it is widely assumed that any HNL, with mass less than the $Z$ mass, must be sterile, and would have no associated charged lepton. Here the PMNS matrix is extended for just one HNL, but others can be added in the same way. The PMNS matrix for the anti-neutrinos is identical to that for neutrinos under CPT symmetry.

Experimental results tend to be presented as upper limits on $|U_{(l=e,\mu,\tau),4}|^{2}$ (usually at the 95 $\%$ confidence level) as a function of $m_{N}$, the mass of the possible HNL being sought. The probability of a fourth neutrino state interacting with the electron ($|U_{e4}|^{2}$) or muon ($|U_{\mu4}|^{2}$) has tight constraints \cite{PDG2020}, limits on $|U_{\tau 4}|^{2}$ are weaker, motivating the possibility that $|U_{\tau 4}|>>|U_{e 4}|$, $|U_{\mu 4}|$. 

This article presents a summary of current bounds on the couplings to each of the active neutrino states. In Sec.\ref{sec:old} long-established limits on all three mixing strengths are discussed. In Sec.~\ref{sec:new} recent results, which have been published in the last few years, are presented. In Sec.~\ref{sec:babar} the most recent result from ~\babar~, which places new limits on $|U_{\tau 4}|^{2}$ in the 100  MeV/$c^{2}$ $< m_{4} <$ 1300  MeV/$c^{2}$ mass range. Existing bounds in the range $\sim$ 300  MeV/$c^{2}$ to $\sim$ 1  GeV/$c^{2}$ range are particularly weak. In Sec.~\ref{sec:future} future projections for near-term projects are discussed and compared to current limits.

\section{\label{sec:old} Experimental Searches}

There are several means of searching for HNLs, depending on the mass regime and proposed coupling to the active neutrinos. Additional neutral leptons can be searched for in cosmology, in colliders, or in high-intensity experiments. Sterile neutrinos can be responsible for contributions to electric and magnetic leptonic dipole moments, and facilitate many rare transitions and decays. Consequently, searches for additional neutrinos cross all frontiers of particle physics.

    \subsection{Established Limits, prior to 2020}

\begin{figure}[t]
     \centering
     \begin{subfigure}[b]{0.5\textwidth}
         \centering
         \includegraphics[width=\textwidth]{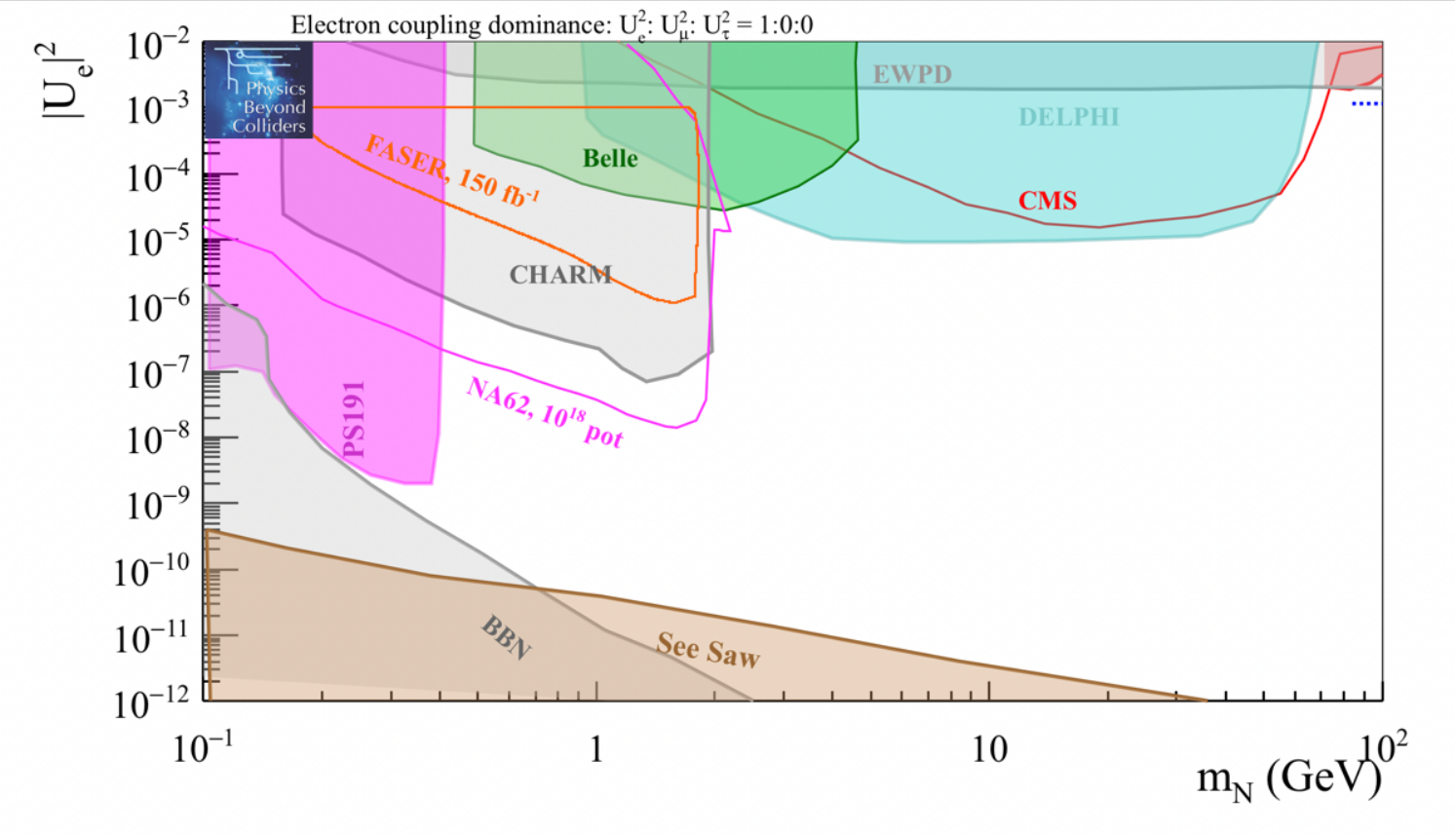}

     \end{subfigure}
     \hfill
     \begin{subfigure}[b]{0.5\textwidth}
         \centering
         \includegraphics[width=\textwidth]{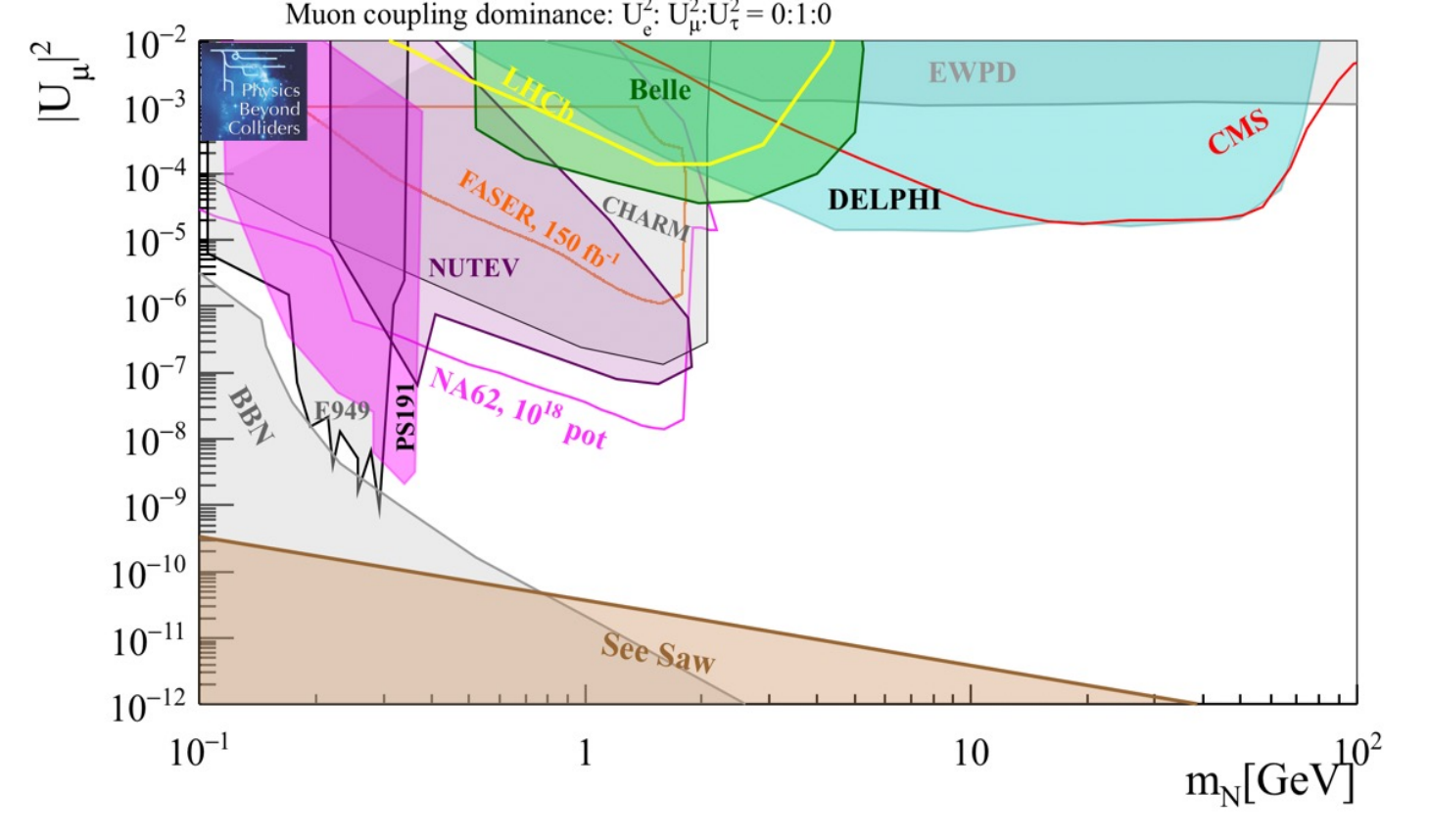}

     \end{subfigure}
     \begin{subfigure}[b]{0.5\textwidth}
         \centering
         \includegraphics[width=\textwidth]{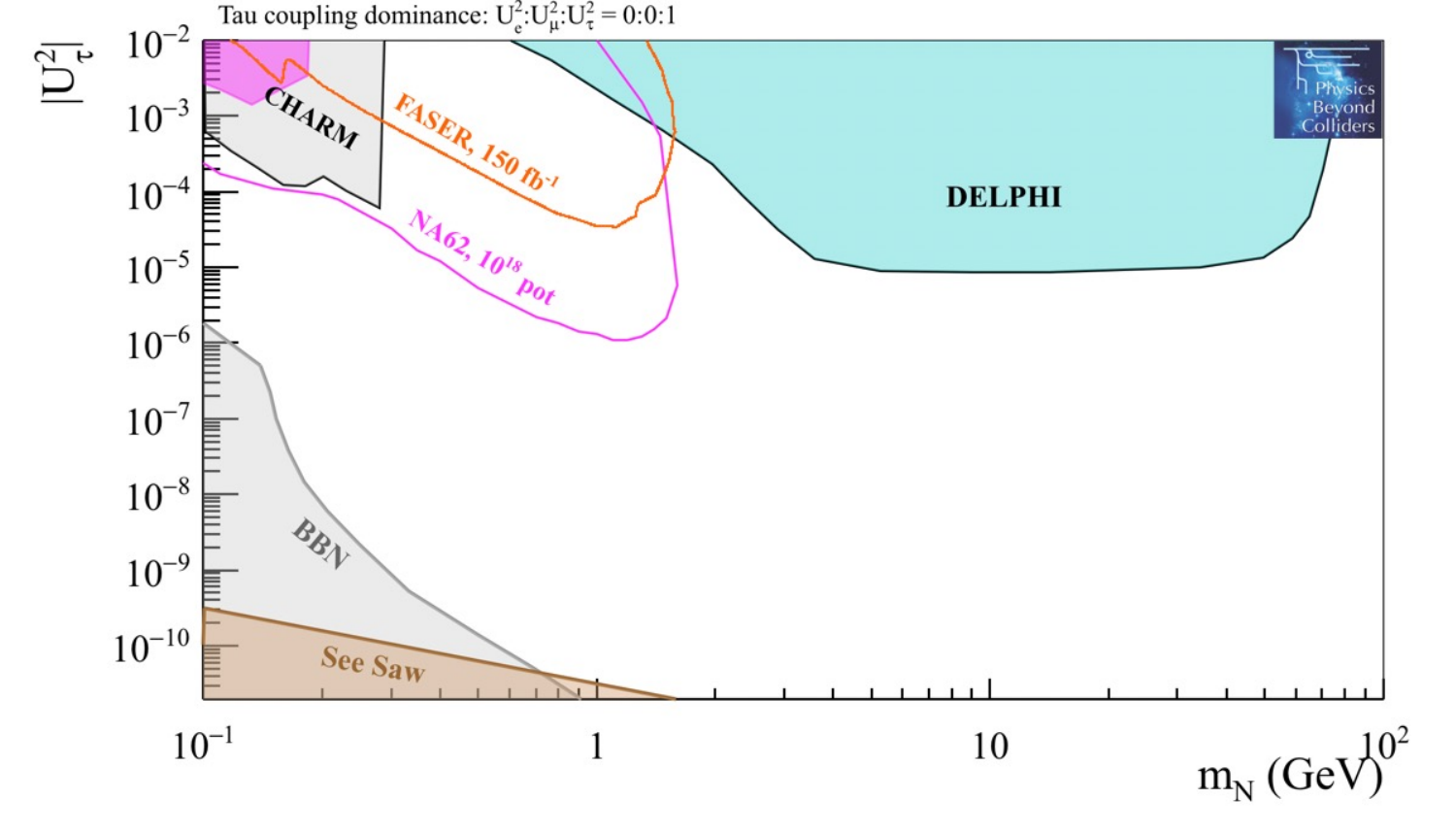}

     \end{subfigure}
     \caption{Sensitivity to Heavy Neutral Leptons with coupling to the (top) electron, (middle) muon and (bottom) tau lepton. Current bounds (filled areas) and near (next 5 years) future physics reach. All plots taken from Ref. \cite{PBC}.}
              \label{fig:limits_tau}
\end{figure}

\subsubsection{Limits on Mixing with Electron and Muon Neutrinos}
Mixing of heavy neutrinos with both $\nu_{e}$ and $\nu_{\mu}$ can be probed by searching for bumps in the missing-mass distribution of pions and kaons leptonic decays. Existing bounds come from a range of experiments: 

\begin{itemize}
    \item \textbf{PS 191 at CERN:} \cite{PS191} designed to search
for neutrino decays in a low-energy neutrino beam. No such events were found and limits were placed. 

\item \textbf{CHARM at CERN:} \cite{Capone:1985vt} which conducted a search for HNLs in a prompt neutrino beam produced by dumping 400 GeV protons in a Cu target. Visible decays with electrons or muon in the final state: $e^{+}e^{-}\nu_{e}, \mu^{+}e^{-}\nu_{e},e^{+}\mu^{-}\nu_{\mu}$ and $\mu^{+}\mu^{-}\nu_{\mu}$ were sought. This search provided limits on $|U_{e4}|^{2},|U_{\mu 4}|^{2}<10^{-7}$ for masses up to 1.5 GeV/$c^{2}$. HNLs were also sought through assuming a neutral-current neutrino interaction in the CHARM calorimeter, and looking for neutrinos decaying to muons and hadrons. This search was sensitive to masses of 0.5-2.8GeV/$c^{2}$ and provided limits of $|U_{\mu 4}|^{2}<10^{-4}$.

\item \textbf{Belle at KEK:} \cite{Belle:2013ytx} used 772 million $B\bar{B}$ pairs
and their leptonic and semi-leptonic decays mesons decays, $B \rightarrow (X) l \nu_{R}$, where $l = e, \mu$ and $X$ was the semi-leptonic case and either a charmed meson $D$($D^{*}$), a light meson (e.g. $\pi.\rho,\eta$). This search had sensitivity to $|U_{e4}|^{2}$ and $|U_{\mu 4}|^{2}$ in a range of masses between the kaon and the $B$ meson mass.

\item \textbf{DELPHI at LEP:} \cite{DELPHI:1996qcc} the best limits on all three mixing strengths in the mass range above the $B$ meson mass come from the DELPHI detector at LEP. There $3.3 \times 10^{6}$ hadronic $Z^{0}$ decays were analyzed. Four separate searches were performed for short-lived HNL production resulting in jets and for longer living HNLs giving detectable secondary vertices or calorimeter clusters. An upper limit for the
branching ratio $BR(Z^{0}\rightarrow \text{HNL} + \bar{\nu})$ of about $1.3\times 10^{-6}$ at 95$\%$ confdence level was found for masses between 3.5 and 50 GeV/$c^{2}$.

\item \textbf{CMS at LHC:} \cite{CMS:2018iaf} used their three prompt charged lepton samples in any
combination of electrons and muons collected at a center-of-mass energy of 13 TeV. Their first results, corresponding to an integrated luminosity of 35.9$fb^{-1}$. The search is performed in the HNL mass range between 1 GeV and 1.2 TeV.
\end{itemize}

The allowed range of couplings is bounded from below
by the Big Bang Nucleo-synthesis (BBN) constraint \cite{Ruchayskiy:2012si}, 
which ensures HNLs wouldn't live long enough in the early Universe to cause an an overabundance of Helium-4, and the See-saw limit \cite{Canetti:2010aw}, below which the mixing to active neutrinos is too weak to produce the observed oscillations.

\subsubsection{Limits on Mixing with Muon Neutrinos}

In addition to these searches, there are a few additional experiments which have helped to constrain the ($|U_{\mu4}|^{2}$, $m_{N}$) parameter space:

\begin{enumerate}
    \item \textbf{NuTeV at Fermilab:} \cite{NuTeV} used $2 \times 10^{18}$ 800 GeV protons interacting with a beryllium oxide target and a proton dump, to search for HNLs in the 0.25–2.0 GeV/$c^{2}$ decaying to muonic final states ($\mu \mu \nu, \mu e \nu, \mu \pi$ and $\mu \rho$).  Upper limits were placed on $|U_{\mu 4}|^{2}$  down to $\mathcal{O}(10^{-7})$.

    \item \textbf{E949 at BNL:} \cite{E949} searched for HNLs from the process $K^{+} \rightarrow \mu^{+} \mu_{R}$ using $1.7 \times 10^{12}$ stopped kaons. They set limits on $|U_{\mu 4}|^{2}$ of the level $\mathcal{O}(10^{-9} - 10^{-7})$ for HNLs of masses 175-300 MeV/$c^{2}$.
\end{enumerate}

\subsubsection{Limits on Mixing with Tau Neutrinos}

Constraints on the ($|U_{\tau4}|^{2}$, $m_{N}$) parameter space are much weaker; the existing constraints come from DELPHI and two other experiments in the low mass region:

\begin{enumerate}
    \item \textbf{CHARM at CERN}: \cite{CHARM}  used the neutrino flux from $2 \times 10^{18}$ 400 GeV protons on a solid copper target to place limits in the 10-290 MeV$/c^{2}$ mass range by re-interpreting the null result of a search for events produced by the decay
of neutral particles into two electrons.

    \item \textbf{NOMAD at  CERN}: \cite{Nomad} used $4.1 \times 10^{19}$ 450 GeV protons on target collected at the WANF facility at CERN. The process $D_{s} \rightarrow \tau \nu_{R}$ followed by the decay $\nu_{R} \rightarrow \nu_{\tau}e^{+}e^{-}$  were analyzed and limits were placed in the 10 to 190 MeV$/c^{2}$ mass range.
\end{enumerate}

    \section{\label{sec:new}New Results, published 2020 - 2022}
    
    In the past few years several new experimental results for couplings with muons and electrons have been published:

\subsection{MicroBooNe: Sterile neutrinos $\mathcal{O}(eV/c^{2})$}

Recent results from MicroBooNe are detailed in Ref. \cite{micro}. MicroBooNE has developed 3 distinct  $\nu_{e}$ searches targeting the MiniBooNE excess:

\begin{enumerate}
    \item an exclusive search for two-body  $\nu_{e}$ charged current quasi-elastic (CCQE) scattering; 
\item a semi-inclusive search for pion-less $\nu_{e}$  events; 
\item an inclusive  $\nu_{e}$ search containing any hadronic final state. 
\end{enumerate}

The combined results were consistent with nominal electron neutrino rate expectations; no excess of electron neutrino events was observed. More results follow.

\subsection{Kaon Searches: Couplings of HNLs $\mathcal{O}(MeV/c^{2})$ mixing with muons and electrons}

The NA62 experiment at CERN recently presented a search for for $K^{+} \rightarrow \mu^{+} + \text{HNL}$, using the 2016-2018 data set in Ref.~\cite{NA62:2021bji}. The analysis found limits of $\mathcal{O}(10^{-8}$) of the neutrino mixing parameter $|U_{\mu 4}|^{2}$ for HNL masses in the range 200-384 MeV/$c^{2}$, with lifetime exceeding 50 ns.

\subsection{LHC Searches: Couplings of HNLs $\mathcal{O}(GeV/c^{2})$ mixing with muons and electrons}

The ATLAS and CMS detectors at CERN have conducted searches for HNLs, detailed in Refs.~\cite{ATLAS:2019kpx,CMS:2022fut}. 

CMS performed a search for HNLs produced with displaced vertices using final states with three charged leptons (electrons or muons), the idea being that
HNLs could be produced through mixing with SM neutrinos. The decay length of these particles can be large enough so that the secondary
vertex of the HNL decay can be resolved with the CMS silicon tracker. The
selected final state would consist of one lepton emerging from the primary proton-proton
collision vertex, along with two leptons forming a displaced, secondary vertex. In this most recent analysis, data totalling 136 fb$^{-1}$ were analyzed. Improved limits on $|U_{e4}|^{2}$ and $|U_{\mu 4}|^{2}$ down to $\mathcal{O}(10^{-7})$ were presented in the 1-20 GeV/$c^{2}$ mass range.

ATLAS analyzed leptonic decays of $W$ bosons extracted using 32.9 - 36.1 fb$^{-1}$ of 13 TeV proton-proton collisions. HNLs could produced through mixing with muon or electron neutrinos. They looked for both prompt and displaced leptonic decay signatures, where the prompt signature requires three leptons produced at the interaction point and the displaced signature comprises a prompt muon from the $W$ boson decay and the requirement of a displaced di-lepton vertex. This search placed limits on $|U_{e4}|^{2}$ and $|U_{\mu 4}|^{2}$ down to $\mathcal{O}(10^{-5})$  for HNL masses in the range 4.5–50 GeV/$c^{2}$.

\section{\label{sec:babar} Couplings of HNLs $\mathcal{O}(MeV/c^{2}-GeV/c^{2})$ mixing with taus at \babar}

The latest analysis from ~\babar~ \cite{MyPaper} presents new limits on  $|U_{\tau4}|^{2}$ in the 100 $<m_{4}<$ 1300 MeV/$c^{2}$ mass range. An overview of the ~\babar~ detector can be found in Ref. ~\cite{BaBar}. The data sample used in this analysis corresponds to an integrated luminosity of 424 fb$^{-1}$.  The average cross-section for $\tau$-pair production of electron-positron annihilation is $ \sigma (e^{+}e^{-} \rightarrow \tau^{+} \tau^{-}) =(0.919 \pm 0.003) $ nb \cite{PDG2020}; so the data sample corresponds to $\sim 4 \times 10^{8}$ produced $\tau$-pairs, before applying any reconstruction or selection criteria.

\subsection{Experimental Strategy}
\label{exp_strat}

This analysis used the approach proposed in Ref. \cite{proposal}, the key principle being that if a HNL is produced in $\tau$ decay, the kinematics
of the visible particles would be modified with respect to SM $\tau$ decay with a massless neutrino.

\begin{figure*}[t]
     \centering
         \includegraphics[width=0.45\textwidth]{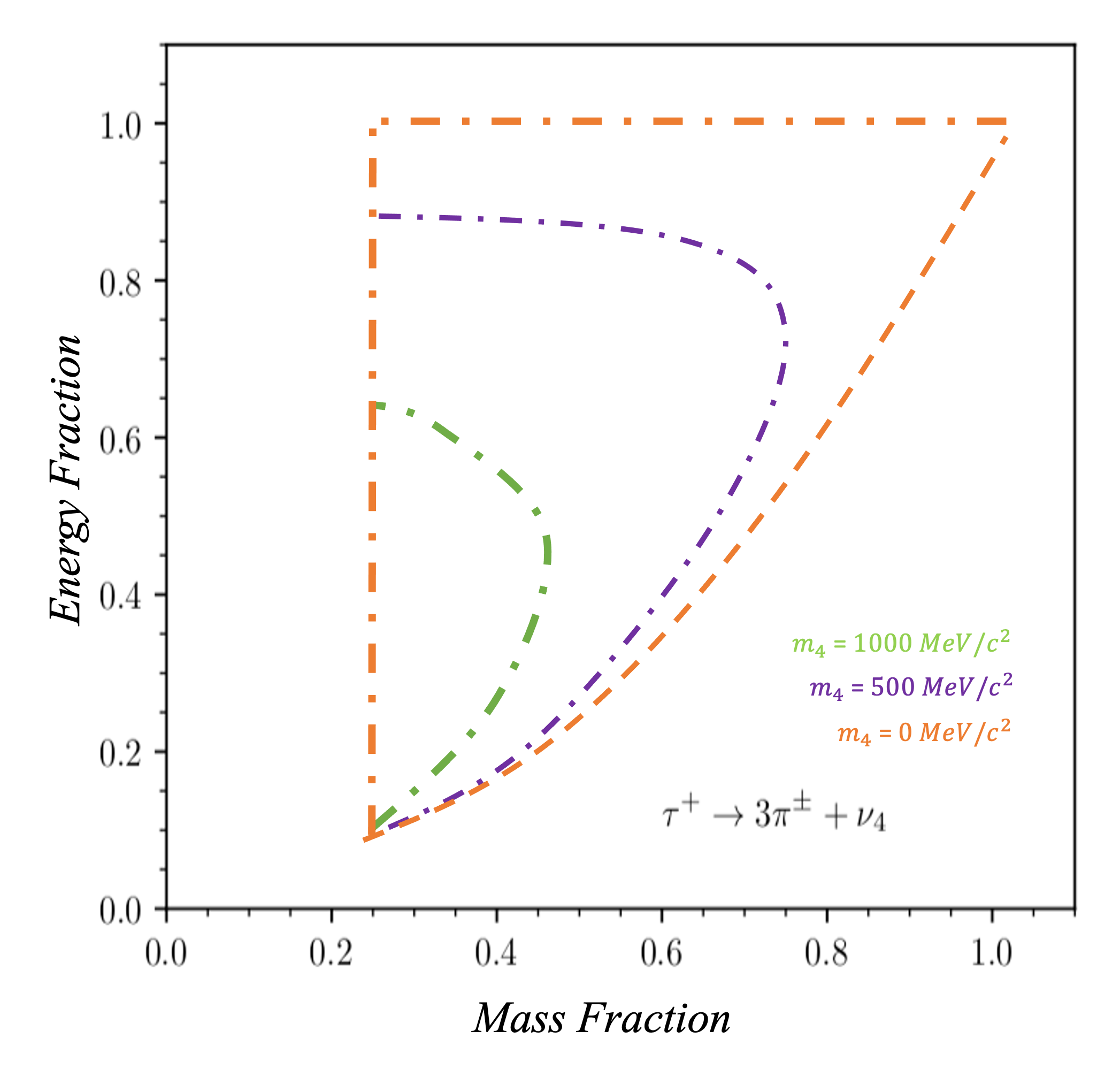}
        \caption{Energy and invariant mass of the hadronic system as fractions of that of the incoming $\tau$ for cases: $m_{4}$ = 0, 500, 1000 MeV/$c^{2}$ are shown. }
        \label{fig:masses}
\end{figure*}

This search studies the 3-prong, pionic $\tau$ decay, which gives access to the region 300$<m_{4}<$1360 MeV$/c^{2}$, where current limits are less stringent. Denoting the three charged pions as a hadronic system $h^{-}$, the decay can be considered a two-bodied:

\begin{equation}  
\tau^{-} \rightarrow h^{-} (E_{h}, \vec{p}_{h}) + \nu  (E_{\nu}, \vec{p}_{\nu}), 
\end{equation}
 where $\nu$ describes the outgoing neutrino state. An analogous equation could be written for the $\tau^{+}$ channel. The allowed phase space of the reconstructed energy, $E_{h}$, and invariant mass, $m_{h}$, of the hadronic system would vary as functions of the mass of the neutrino. As the HNL gets heavier the proportion of the original $\tau$-lepton's energy going to the visible pions diminishes. A visualization is presented in Fig.~\ref{fig:masses}.

In the center-of-mass frame the $\tau$-lepton energy is assumed to be $\sqrt{s}/2$, when there is no initial state radiation. Since the direction of the decaying $\tau$-lepton is not known, we cannot compute the neutrino mass directly, but we know that $E_{h}$ must fall between two extremes that define the kinematically allowed values:

\begin{equation} E_{\tau} - \sqrt{m^{2}_{4} + q_{+}^{2}} < E_{h} < E_{\tau} - \sqrt{m^{2}_{4} + q_{-}^{2}}, \end{equation}
\\
where
\begin{widetext}
\begin{equation}
q_{\pm} = \frac{m_{\tau}}{2} \bigg ( \frac{m_{h}^{2} - m_{\tau}^{2} - m_{4}^{2}}{m_{\tau}^{2}} \bigg ) \sqrt{\frac{E_{\tau}^{2}}{m_{\tau}^{2}} - 1} \pm \frac{E_{\tau}}{2} \sqrt{\big ( 1- \frac{(m_{h}+ m_{4})^{2}}{m_{\tau}^{2}}\big ) \big ( 1- \frac{(m_{h} - m_{4})^{2}}{m_{\tau}^{2}} \big )};
\end{equation}
\end{widetext}
and $3m_{\pi^{\pm}} < m_{h} < m_{\tau} - m_{4}$. As the HNL mass increases, the allowed phase space of the visible system is reduced in the $E_{h}$,$m_{h}$ plane.

A HNL signal is sought by comparing the observed event yield density in the ($m_{h},E_{h}$) plane to a set of template 2D histogram distributions for the background, obtained by simulating all $\tau$ known decays as well as non-$\tau$ background events, and the potential HNL signal for different $m_{4}$ mass values. Although the invariant mass and outgoing hadronic energy ($m_{h}$ and $E_{h}$) are correlated, more information can be extracted by considering both variables simultaneously.  

\subsection{Signal and Background Simulations}

Three potential sources background are considered:

\begin{enumerate}
    \item  $\tau^{-} \rightarrow \pi^{-}\pi^{-}\pi^{+}\nu_{\tau}$, with an outgoing SM neutrino;
    \item other SM $\tau$ decays that have been misidentified as the 3-prong (3 charged pion) decay;
    \item non-$\tau$ backgrounds that have been misidentified as the 3-prong decay.
\end{enumerate}

All SM background yields are estimated from Monte Carlo (MC) simulations which are passed through the same reconstruction and digitization routines as the data. 
All $\tau$-pair events are simulated using the KK2F \cite{kk2f} generator and TAUOLA \cite{tauola} which uses the averaged experimentally measured $\tau$ branching rates as listed in Ref.~\cite{PDG2020}.

Several non-$\tau$ backgrounds are also studied, including:

\begin{itemize}
    \item  $e^{+}e^{-} \rightarrow \Upsilon(4S) \rightarrow B^{+} B^{-}$ and  $e^{+}e^{-} \rightarrow \Upsilon(4S) \rightarrow B^{0}$ $\bar{B}^{0}$ which are simulated using EvtGen \cite{EvtGen};
    \item  $e^{+}e^{-} \rightarrow  u\bar{u} ,  d\bar{d},s\bar{s}$ and   $e^{+}e^{-} \rightarrow  c\bar{c}$ which are simulated using JETSET \cite{ref2} \cite{ref3};
    \item $e^{+}e^{-} \rightarrow \mu^{+} \mu^{-} (\gamma)$ which are simulated using KK2F \cite{ref1}.
\end{itemize} 
A total of 26 signal samples were simulated, one for each of the HNL masses across the range 100 MeV/$C^{2}$ $< m_{4} <$ 1300 MeV/$C^{2}$, at 100 MeV/$C^{2}$~ increments. For each of these HNL masses, both a $\tau^{+}$ and $\tau^{-}$ signal channel were simulated. Signal samples were produced within the \babar~ software environment using KK2F and TAUOLA by changing the value of the outgoing neutrino mass in TAUOLA. The generated signal was passed through the same digitization and reconstruction model as the SM background and data samples. 

\subsection{Analysis Procedure}

A binned likelihood approach is taken in which it is assumed that the contents of a given bin, $i,j$, in the $(m_{h},E_{h})$ data histogram are distributed as a Poisson distribution and may contain events emanating from any of the SM background process, and potentially HNL signal events.  The likelihood to observe the selected candidates in all the $(m_{h}, E_{h})$  bins is the product of the Poisson probability to observe the selected events in each bin:

\begin{widetext}

\begin{align*}
\mathcal{L} =   \prod_{\text{charge}}^{+ -} \bigg ( \prod_{\text{channel}}^{e \mu } \bigg (  \prod^{ij}_{\text{bin}}  \bigg ( \frac{1}{n_{\text{obs},ij}!}\bigg[ N_{\tau,\text{gen}}\cdot |U_{\tau 4}|^{2} \cdot p_{\text{\tiny HNL}, ij}+N_{\tau,\text{gen}}\cdot (1-|U_{\tau 4}|^{2}) \cdot p_{\tau-\text{SM}, ij} + n^{\text{reco}}_{BKG,ij}\bigg]^{(n_{\text{obs}})_{ij}} \times
\end{align*}
\begin{equation}
exp\bigg[-(N_{\tau,\text{gen}}\cdot |U_{\tau 4}|^{2} \cdot p_{HNL, ij} + N_{\tau,\text{gen}}\cdot (1-|U_{\tau 4}|^{2}) \cdot p_{\tau-SM, ij}+ n^{\text{reco}}_{BKG,ij}) \bigg] \bigg )_{\text{bin}}
\times  \prod_{k} f(\theta_{k},\tilde{\theta}_{k}) \bigg )_{\text{channel}} \bigg )_{\text{charge}},
\label{eq:lik}
\end{equation}
\end{widetext}
where $n_{\text{obs}}$ is the number of observed events in the bin $ij$, $N_{\tau,\text{gen}}$ is the number of generated $\tau$'s, $p_{HNL (\tau-SM), ij}$ is the probability of a reconstructed event being in a given bin in the HNL ($\tau-SM$) 2D template and $n^{\text{reco}}_{BKG,ij}$ is the expected number of non-$\tau$ background events. The final product is a set of Gaussian nuisance parameters. The expression involves a product over all bins, $ij$, over the two 1-prong channels, and over both $\tau$-lepton charges ($\pm$). 

A test statistic, $q$, can be defined as:

\begin{equation}
    q = -2 \text{ln} \bigg (  \frac{\mathcal{L}_{H_{0}}(|U_{\tau 4}|_{0}^{2};\hat{\hat{\theta}}_{0},\text{data})}{\mathcal{L}_{H_{1}}(|\hat{U}_{\tau 4}|^{2};\hat{\theta},\text{data}) } \bigg ) = -2\text{ln}(\Delta \mathcal{L}),
\end{equation}
where $\mathcal{L}$ in both the numerator and denominator describes the maximized likelihood for two instances. The denominator is the maximized (unconditional) likelihood giving the maximum likelihood estimator of $|U_{\tau 4}|^{2}$ and the set of nuisance parameters ($\hat{\theta}$); $\hat{\theta}$ is a vector of nuisance parameters that maximize the likelihood. In the numerator the nuisance parameters are maximized for a given value of $|U_{\tau 4}|^{2}$, i.e it is the conditional maximum-likelihood. The ratio, $LR$, is consequently a function of $|U_{\tau 4}|^{2}$ through the numerator. It must be noted that the numerator denotes the hypothesis for any given value of $|U_{\tau 4}|^{2}$ (including the background only case i.e. $|U_{\tau 4}|^{2}=0$). Reference \cite{Cowan:2010js} provides more details on likelihood-based tests. The analysis aims to find the value of $|U_{\tau 4}|^{2} $ that minimizes this quantity at the 95 $\%$ confidence level.

\subsection{Uncertainties}

Table~\ref{table:syslist} lists the relative contribution for each of the systematic uncertainties on the normalization. These are parameterized as Gaussian nuisance parameters.

\begin{table*}[t!]
\centering
\caption{Systematic uncertainty contribution to the event yield (in $\%$)  from each source, based on comparisons between MC simulations and data. }
\begin{tabular}{c |  c } 
 \hline 
Uncertainty &  Yield Change ($\pm$)\\ [0.5ex] 
 \hline
 \hline
Luminosity &  $0.44 \%$\\
$\sigma(ee \rightarrow \tau \tau)$ &$0.31 \%$\\
\hline
Branching Fractions (1 prong)  & e: 0.22$\%$ \\
&$\mu$: 0.22$\%$\\
\hline
Branching Fractions (3 prong)  & 3$\pi$: 0.57$\%$ \\
\hline
PID Efficiency & $e: 2\%$ \\
& $\mu $: 1$\%$ \\
&  $\pi $:  3$\%$ \\
   \hline
Bhabha Contamination &  0.2$\%$\\\
$q\bar{q}$ Contamination (data) &0.1$\%$\\
\hline
   Tracking Efficiency & negligible\\
   Detector Modeling & negligible \\
\hline
Beam Energy & negligible\\
Tau Mass & negligible\\
\hline
\end{tabular}
\label{table:syslist}
\end{table*}

In addition to these yield uncertainties, which effect all bins uniformly, ``shape" uncertainties must also be accounted for. These come from inefficiency in the  MC modelling. For many hadronic $\tau$ decay channels the relative uncertainties from experimental results are large.

A $\tau$-lepton decay to three charged pions is mediated by the  $a_{1}(1260)$ resonance which decays through the intermediate $\rho \pi$ state. In the MC samples used in this analysis the PDG \cite{PDG2020} average of $m_{a_{1}} = 1230 \pm 40$ MeV/$c^{2}$ and a Breit-Wigner averaged width of $\Gamma_{a_{1}} = 420 \pm 35 $ are used; Ref.~\cite{PDG2020}  quotes the estimated width to be between 250 - 600. The uncertainty associated with the $a_{1}$ resonance represents the dominant contribution to the systematic error in our measurement. In order to understand the effects of the uncertainty on the $a_{1}$ mass on the final results in this analysis several additional MC simulations were built, in which the $m_{a_{1}}$ was varied to $\pm 1 \sigma$ of the experimental average (where $\sigma = 40$ MeV/$c^{2}$ ).

\subsection{Results}

\begin{figure*}[t]
         \centering
         \includegraphics[width=5in]{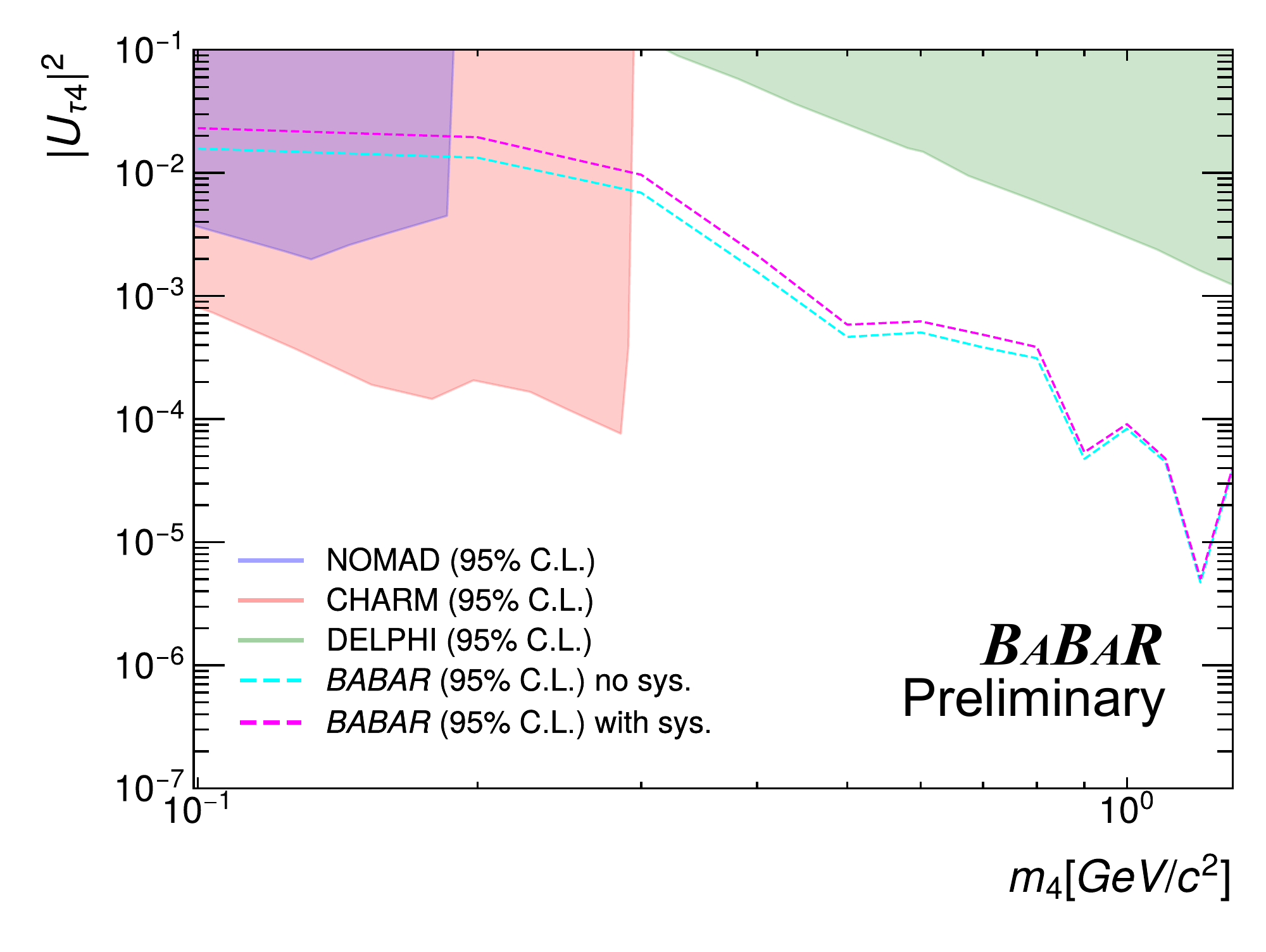}
        \caption{Upper limits at 95$\%$ C.L. on $|U_{\tau 4}|^{2}$. The magenta line represents the result when uncertainties are included. The magenta line is expected to be a very conservative upper limit. }
        \label{fig:limits}
\end{figure*}

Figure~\ref{fig:limits} shows the upper limit at the 95$\%$ confidence level provided by this analysis using the described binned likelihood technique. The magenta line represents the upper limit when all systematic uncertainties are considered. To characterize deviations due to the uncertainty on $\Gamma_{a_{1}}$ the more conservative PDG estimates are used. The dominant systematic uncertainty is, by far, that due to the assumptions made within our simulation, the main contribution being uncertainty in the intermediate resonances for the $\tau$ 3-prong channel, and the dominant $\tau$ backgrounds. The relative systematic uncertainty decreases as the mass of the hypothetical HNL increases, this is expected since the effects of the modeling uncertainty become less apparent at higher HNL masses.

    \section{\label{sec:future}Future projections}
    
    References \cite{PBC,Abdullahi:2022jlv} provide comprehensive reviews of projections for near and far term projects. Figure-\ref{fig:limits_tau} show near-term projections (unfilled lines) for two experiments expected to publish data in the next few years. The new limit from ~\babar~ is competitive with both the projected limits from FASER (at 150 fb$^{-1}$) and NA62 (at $10^{18}$) POT. The ~\babar~ technique is also applicable to data at Belle-II and elsewhere.
    
    There are several other experiments planned for the next decade that will also improve limits on mixing to all three active neutrinos. The SHiP experiment \cite{Gorbunov:2020rjx} aims to make improvements in all three parameter spaces, with projected limits down to $10^{-9}$ in the electron and muon sector and $10^{-7}$ in the tau sector, in the few GeV/$c^{2}$ region. Other limits are expected from DUNE \cite{Carbajal:2022zlp}, CODEX-b \cite{PhysRevD.97.015023} and MATHUSLA \cite{MATHUSLA:2019qpy}.
    
    Looking much further ahead, the FCC-ee could provide very powerful limits, down to $\mathcal{O}(10^{-9}-10^{-12})$ \cite{FCCee,Shen:2022ffi} for HNL  masses of 5-80 GeV/$c^{2}$.
    
    \section{ Conclusions}
To conclude, the existence of Heavy Neutral Leptons can provide solutions to many issues which exist within the Standard Model. Depending on the mass and coupling of the HNL to active neutrinos these new particles can be produced in a range of channels. Consequently the experimental program searching for them extends well beyond the neutrino sector. This article has documented recent results from searches at neutrino experiments, beam-dumps and collider-based experiments. This includes a new upper limit on $|U_{\tau 4}|^{2}$ set by ~\babar. The technique presented in Sec.~\ref{sec:babar} can also be applied future searches at Belle-II. the results presented are competitive with projections for experiments coming online in the next few years. 

The next decade brings with it new high-intensity searches for HNLs coupling to all three active neutrinos, many experiments are able to access mass-scales which are predicted by well-motivated models. There is no denying that we sit on the precipice of a very interesting time in the search for these elusive, beyond SM particles.

\begin{acknowledgments}

We are grateful for the extraordinary contributions of our PEP-II colleagues in achieving the excellent luminosity and machine conditions that have made this work possible. The success of this project also relies critically on the expertise and dedication of the computing organizations that support \babar. The collaborating institutions wish to thank SLAC for its support and the kind hospitality extended to them. This work is supported by the US Department of Energy and National Science Foundation, the Natural Sciences and Engineer- ing Research Council (Canada), Institute of High Energy Physics(China),the Commissariat al' Energie Atomique and Institut National de Physique Nucleaire et de Physique des Particules (France), the Bundesministerium fur Bildung und Forschung and Deutsche Forschungsge meinschaft (Germany), the Istituto Nazionale di Fisica Nucleare (Italy), the Foundation for Fundamental Research on Matter (The Netherlands), the Research Council of Norway, the Ministry of Science and Technology of the Russian Federation, and the Particle Physics and Astronomy Research Council (United Kingdom). Individuals have received support from CONACYT (Mexico), the A. P. Sloan Foundation, the Research Corporation, and the Alexander von Humboldt Foundation.

\end{acknowledgments}
    
\bigskip 
\bibliography{paper}

\end{document}